\documentclass[conference]{IEEEtran}
\IEEEoverridecommandlockouts
\usepackage{cite}
\usepackage{amsmath,amssymb,amsfonts}
\usepackage{algorithmic}
\usepackage{float}
\usepackage{graphicx}
\usepackage{textcomp}
\usepackage{xcolor}
\def\BibTeX{{\rm B\kern-.05em{\sc i\kern-.025em b}\kern-.08em
    T\kern-.1667em\lower.7ex\hbox{E}\kern-.125emX}}
\begin{document}

\title{Multi-dimensional Energy Limitation in Sphere Shaping for Nonlinear Interference Noise Mitigation}

\makeatletter
\newcommand{\linebreakand}{%
  \end{@IEEEauthorhalign}
  \hfill\mbox{}\par
  \mbox{}\hfill\begin{@IEEEauthorhalign}
}
\makeatother

\author{
\IEEEauthorblockN{Jingtian Liu}
\IEEEauthorblockA{\textit{Télécom Paris, Institut Polytechnique de Paris}\\
19 place Marguerite Perey, 91120 Palaiseau, FRANCE \\
jingtian.liu@telecom-paris.fr}
\and
\IEEEauthorblockN{Élie Awwad}
\IEEEauthorblockA{\textit{Télécom Paris, Institut Polytechnique de Paris}\\
19 place Marguerite Perey, 91120 Palaiseau, FRANCE \\
elie.awwad@telecom-paris.fr}
\linebreakand
\IEEEauthorblockN{Yves Jaouën}\centering
\IEEEauthorblockA{\textit{Télécom Paris, Institut Polytechnique de Paris}\\
19 place Marguerite Perey, 91120 Palaiseau, FRANCE \\
yves.jaouen@telecom-paris.fr}
}

\maketitle

\begin{abstract} We propose Four-Dimensional (4D) energy limit enumerative sphere shaping (ESS) of $M$-QAM signaling to minimize rate loss and improve the transmission performance over non-linear WDM optical-fiber systems. Simulation results show that the proposed scheme outperforms the conventional ESS by $0.19$~bit/4D-symbol in achievable information rate over a $205$-km single-span link and a WDM transmission of five polarization-division-multiplexed channels with $400$-Gbit/s net rate per channel. We also study the achieved performance over several shaping block lengths and show that the achieved gains do not scale well over multi-span systems. 
\end{abstract}


\section{Introduction}
Probabilistic amplitude shaping (PAS) has demonstrated excellent performance over linear AWGN channels. PAS can be implemented through several approaches, among which we cite enumerative sphere shaping (ESS)~\cite{gultekin2019enumerative} and constant composition~\cite{schulte2015constant} distribution matching (CCDM). When it comes to dealing with non-linear interference (NLI) in optical fiber transmission, PAS may be more susceptible to performance degradation~\cite{fehenberger2019analysis}. Recently, several works aimed at improving the non-linear performance of PAS schemes by limiting the 1-dimensional (1D) amplitude variance or kurtosis of ESS schemes~\cite{gultekin2021kurtosis,gultekin2022mitigating} or by limiting the 2D energy dispersion index (EDI) of CCDM schemes~\cite{wu2022list}. Inspired by these advancements, we propose a new distribution matching approach to alleviate the distortions introduced by Kerr effects in optical fiber systems. Our design targets lowering 4D energy variations where the four dimensions refer to the in-phase and quadrature parts of two orthogonal polarization tributaries of the transmitted signal. The proposed scheme is based on polarization-division-multiplexed (PDM) quadrature-and-amplituded (QAM) signaling along with ESS for amplitude shaping. ESS is chosen for its reduced rate loss when it is applied over short shaping block lengths compared to CCDM~\cite{gultekin2019enumerative}. 

In this paper, we propose a 4D energy limit over ESS shaping to achieve either an improved non-linear tolerance or a lower rate loss than state-of-the-art kurtosis-ESS (K-ESS)~\cite{gultekin2021kurtosis} and band-limited ESS (B-ESS)~\cite{gultekin2022mitigating}, and this is done by tuning the design parameters of the new scheme. Simulation results show that the proposed shaping outperforms conventional ESS in two different scenarios when applied over $64$-QAM. We measure a gain of $0.19$~bit per 4D-symbol in the achievable information rate (AIR) over a $205$-km single-span standard single-mode-fiber (SSMF) link and a WDM transmission of five channels with a net rate of $400$~Gbit/s per wavelegnth at $50$~GBaud and a block length of $N=108$. Smaller gains are observed for larger block lengths over a $5$-span $400$-km link and a WDM transmission of five $110$~GBaud channels.

\section{Principle}
ESS can be generated and applied over $1$, $2$, $4$ or even more dimensions. In Fig.~3 from~\cite{skvortcov2021huffman}, the authors show an illustration of 1D, 2D, and 4D symbol mapping strategies of ESS schemes. For 1D mapping, 4D symbols are shaped using $4$ independent amplitude sequences (one ESS per dimension). For 2D mapping, each polarization, namely $X$ and $Y$, uses one ESS-shaped amplitude sequence. For 4D mapping, a single-shaped sequence is used for a wavelength channel. The latter mapping offers the advantage of reducing the probability of having equal amplitudes at the same time slot; hence, high peak power values appear less frequently and lead to lower non-linear distortions. This advantage was demonstrated in several works~\cite{skvortcov2021huffman,gultekin2021kurtosis,fehenberger2019analysis}.

\begin{figure*}[t]
    \centering
    \includegraphics[width=1\linewidth]
    {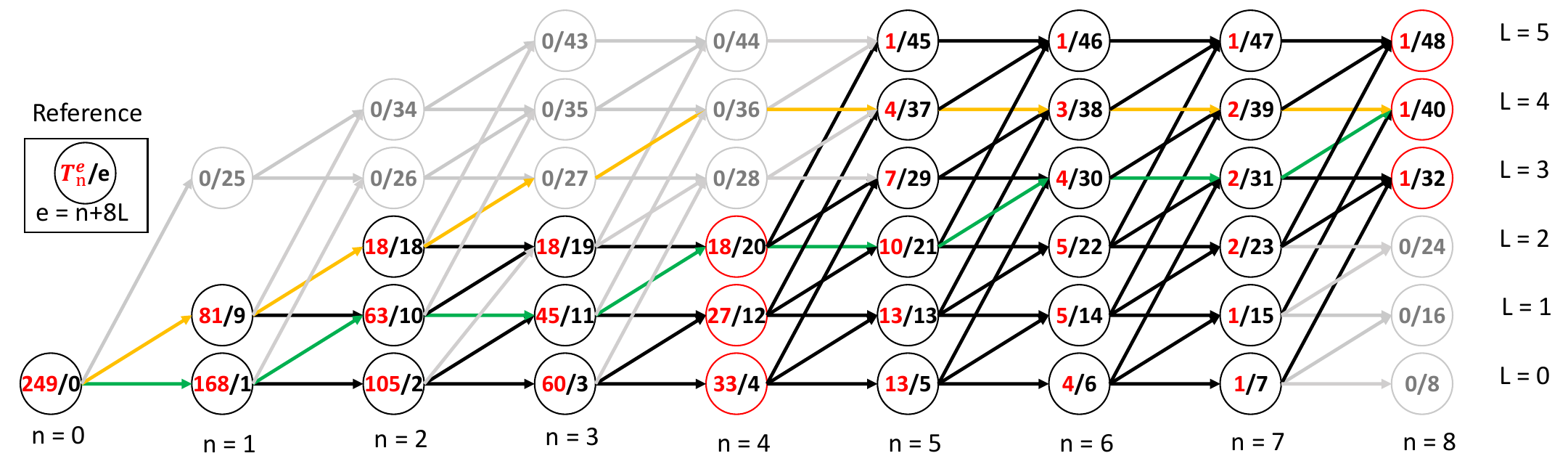}
    \caption{4D-Band-limited-trellis using four amplitudes $\left\{ 1 ,3 , 5 , 7 \right\}$, $E_{\mathrm{max}}=48$, and block length $N=8$.}
    \label{fig:4D-BL-ESS}
\end{figure*}

\begin{table*}[htb]
\large
\centering
\caption{Metrics for five different ESS schemes with $N=108$, rate = $1.5$~bits/1D amplitude}
\label{tab:results}
\begin{tabular}{|c|c|c|c|c|c|c}
\hline
 & ESS & 1D-BL-ESS & 4D-BL-ESS (Linear) & 4D-BL-ESS (Nonlinear) & K-ESS \\  \hline
$E_{\mathrm{max}}$ & 860 & 996 & 948 & 996 & 1156 \\ \hline
 $\overline{E}_{1D}$ & 7.85 & 9.14 & 8.63 & 9.09 & 8.27 \\ \hline
 $\mathrm{\sigma^2_{1D}}$ & 1.75 & 1.49 & 1.52 & 1.45 & 1.16 \\ \hline
$\mathrm{\mu_{4,2D}}$ & 1.87 & 1.71 & 1.69 & 1.64 & 1.57 \\ \hline
%$\mathrm{\sigma^2_{E_{4D-norm}}}$ & 0.44 & 0.27 & 0.28 & 0.23 & 0.28 \\ \hline
%$\mathrm{\mu_{4,E_{4D}}}$ & 4.03 & 3.02 & 2.92 & 2.74 & 3.35 \\ \hline
$\mathrm{EDI}~(W_1=42)$ & 3.13 & 0.65 & 0.72 & 0.48 & 3.34 \\ \hline
$\overline{\mu}_{4,E_{4D}} (W_2=54)$ & 3.63 & 2.8 & 2.84 & 2.69 & 3.17 \\ \hline
$\mathrm{Rate~loss~[bit/1D]}$ & 0.03 & 0.13 & 0.097 & 0.13 & 0.07 \\ \hline
\end{tabular}
\end{table*}

While both K-ESS~\cite{gultekin2021kurtosis} and B-ESS~\cite{gultekin2022mitigating} (that we call 1D-BL-ESS in the following) use 4D mapping, they introduce energy constraints over a single dimension to limit the variations of each 1D amplitude. However, the strength of non-linear effects depends on the variations of energy levels of the 4D symbols in consecutive time slots. Hence, by constraining energy variations of 4D symbols, we can improve the performance and achieve a better trade-off between linear and non-linear gains. The following toy example illustrates this advantage. Consider an ESS with a block length $N=8$ and two amplitude sequences $(3,3,3,3,1,1,1,1)$, highlighted in yellow in Fig.~\ref{fig:4D-BL-ESS}, and $(1,3,1,3,1,3,1,3)$, highlighted in green. These sequences have the same 1D average energy, variance, and kurtosis, and generate the two consecutive 4D-symbols: $[\left(\pm 3\pm 3i,\pm 3\pm 3i\right)^T, \left(\pm 1\pm 1i,\pm 1\pm 1i\right)^T]$ and $[\left(\pm 1\pm 3i,\pm 1\pm 3i\right)^T, \left(\pm 1\pm 3i,\pm 1\pm 3i\right)^T]$ respectively where the first symbol is attributed to the polarization $X$ and the second symbol to the polarization $Y$ in each pair, and $(\cdot)^T$ is the transpose of $(\cdot)$. Symbols from the first sequence show a large temporal 4D energy fluctuation from $36$ to $4$ while symbols from the second one have a constant 4D energy equal to $20$.

In Fig.~\ref{fig:4D-BL-ESS}, we illustrate the new 4D-band-limited ESS (4D-BL-ESS) scheme through a trellis using 64-QAM that has 4 amplitudes in 1D: $\left\{ 1 ,3 , 5 , 7 \right\}$, a block length $N=8$, and a maximum accumulated energy limit $E_{\mathrm{max}}=48$. $n$ and $L$ are respectively the column and row indices of the nodes in the trellis. Each path in the trellis correspond to a possible amplitude sequence. In each node, the number in black $e$ is the intermediate accumulated energy up to that state, and the number in red $T^e_n$ is the number of paths that lead from this state to a final state at $n=N$. The accumulated 1D energy in the $4$-th and $8$-th positions (red nodes) corresponds to the accumulated 4D energy in the $1$-st and $2$-nd time slots respectively. Sequences in grey are deleted based on a 4D-energy constraint instead of a 1D-constraint as initially done in~\cite{gultekin2022mitigating}. For this example, the constraint is: $4+28(i-1)\leq E_{4D,i} \leq 28+28(i-1)$ where $E_{4D,i}$ is the accumulated 4D energy in $i$-th time slot. The construction used in the toy example can be generalized by tuning the constraints on the accumulated energy:
\begin{equation}
    A+iK_1 \leq E_{4D,i} \leq \mathrm{min}(B+iK_2,E_{\mathrm{max}})
    \label{Equation.1}
\end{equation}
where the parameters $A, B, K_1, K_2$ define the upper and lower limits of 4D energies in the trellis. 

\begin{figure*}[htb]
    \centering
    \includegraphics[width=0.92\linewidth]
    {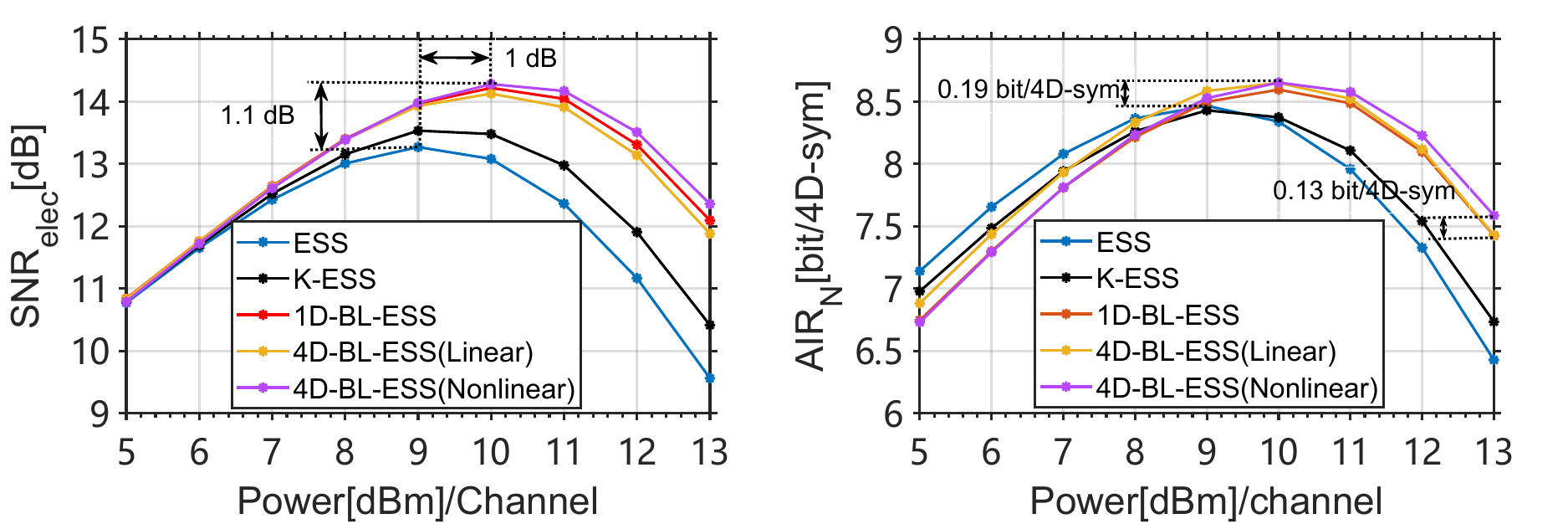}
    \caption{Left: Electrical SNR versus Power per channel in the non-linear channel. Right: GMI versus Power per channel. Simulation of conventional ESS, K-ESS, 1D-BL-ESS, 4D-BL-ESS(Linear), and 4D-BL-ESS(Nonlinear) in $50$~GBaud over $1\times205$~km SSMF link, 5 WDM channels, 55 GHz channel spacing, net rate per channel $400$~Gbit/s.}
    \label{fig:205km_1_span_nonlinear}
\end{figure*}

For ease of comparison between different schemes, we evaluate several metrics related to the linear and non-linear performance using the same signaling parameters as in~\cite{gultekin2021kurtosis, gultekin2022mitigating}, i.e. a shaping rate of $1.5$~bits/amplitude, a block length $N=108$, an LDPC FEC code with a rate of $r_c=5/6$ and acheive a net bit rate of $8$~bits/4D symbol which is calculated as in~\cite{Schmalen18}. For conventional ESS, the net bit rate for each amplitude is determined only by the chosen value of $E_{\mathrm{max}}$. However, for all three band-limited ESS and the K-ESS, we start with a larger energy limit $E_{\mathrm{max}}$, hence with a larger bit rate, and we then reduce the bit rate through constraints such as band limiting the enrergy or limiting the kurtosis until we reach the same net rate as the one of conventional ESS. In~\cite{gultekin2021kurtosis,gultekin2022mitigating}, the average energy of 1D amplitudes $\overline{E}_{1D}$ is used to assess the linear penalty, also the variance of 1D amplitudes $\mathrm{\sigma^2_{1D}}$ and the kurtosis of 2D amplitudes $\mathrm{\mu_{4,2D}}$ are the used metrics to assess the non-linear gain. Hence, in Table~\ref{tab:results}, we compute these metrics for five schemes: the conventional ESS~\cite{gultekin2019enumerative}, the kurtosis-ESS (K-ESS)~\cite{gultekin2021kurtosis}, the 1D-BL-ESS~\cite{gultekin2022mitigating} and two proposed 4D-BL schemes. We also add two 4D metrics: the energy dispersion index (EDI) defined in~\cite{wu2021temporal}, and the windowed kurtosis defined in~\cite{cho2022shaping}. These two 4D metrics were shown to be better suited for assessing non-linear gains of non-iid (independent and identically distributed) signaling schemes such as short block-length PAS  modulations~\cite{cho2022shaping,wu2021temporal,gultekin2022temporal}. The length of the averaging windows $W_1$ for the EDI and $W_2$ for the windowed kurtosis are computed as Eq.~(2) in~\cite{wu2021temporal} and Eq.~(7) in~\cite{cho2022shaping}. Their values in Table~\ref{tab:results} are computed for the first transmission scenario over a single-span $205$-km SSMF link simulated in the next section. In this paper, we design and study two 4D-BL schemes: the first one is optimized for the linear regime, is denoted 4D-BL-ESS (Linear) and has a lower accumulated energy threshold $E_{\mathrm{max}}$ than its 1D-BL counterpart, thus a lower rate loss (linear gain) and almost the same energy variations as 1D-BL-ESS; and the second scheme is optimized for the non-linear regime, is denoted 4D-BL-ESS (Nonlinear) and has the same accumulated energy threshold $E_{\mathrm{max}}$ as 1D-BL-ESS (hence, the same rate loss too) but it has lower 4D energy variations. The rate loss for each scheme is computed and reported in the last row of Table~\ref{tab:results}. We can see that all band-limited ESS schemes show lower energy variations than the conventional ESS (first column), but have a higher rate loss due to the addition of energy constraints that reduce the number of allowed amplitude combinations. At the same net rate, K-ESS has the lowest 1D energy variation, but the highest 4D energy variation, which makes its nonlinear performance worse than the other schemes as will be shown in the next section. 

\begin{figure*}[htb]
    \centering
    \includegraphics[width=1\linewidth]
    {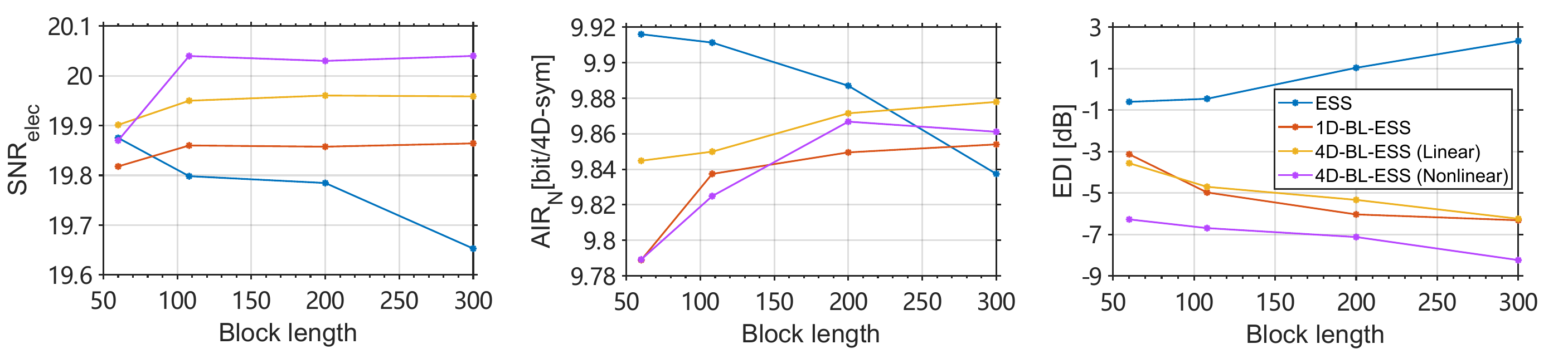}
    \caption{Simulation of conventional ESS, 1D-BL-ESS, 4D-BL-ESS(Linear), and 4D-BL-ESS(Nonlinear) for a $110$~GBaud transmission over $5\times80$~km SSMF link, 5 WDM channels, $112.5$~GHz channel spacing, net rate per channel: $838$~Gbit/s. Electrical SNR (left) and AIR (middle) vs block length at optimal power ($3$~dBm). EDI (right) with $W_1=200$ versus block length.}
    \label{fig:400km_5_span_nonlinear}
\end{figure*}

\section{Simulation setup and results}

\subsection{Single-span $205$-km $50$GBaud transmission}

We evaluate the performance of the two proposed 4D-BL-ESS with $N=108$ under the same conditions as the ones used in~\cite{gultekin2021kurtosis,gultekin2022mitigating}. The transmitted signals are $5$ dual-polarization WDM channels, with a baud rate of $50$~GBaud per channel. The raw data rate is $600$~Gbit/s. For all ESS schemes, we use an LDPC code of length $n=64800$~bits from the DVB-S2 standard with a rate of $r_c=5/6$. The achieved net bit rate is $8$~bits per 4D symbol. The channel spacing is $55$~GHz. Root-raised-cosine pulse shaping is applied with a roll-off factor of $0.1$. The signal propagation is simulated through the split-step Fourier method based on the nonlinear Schrödinger and Manakov equations. SSMF fiber segments are simulated with a chromatic dispersion coefficient $D=17\mathrm{ps/nm/km}$, non-linearity coefficient $\gamma = 1.3\mathrm{(W\cdot km)^{-1}}$, and an attenuation $\alpha =0.2~\mathrm{dB/km}$ at $\lambda=1550~\mathrm{nm}$. Polarization mode dispersion (PMD) is added with $0.04~\mathrm{ ps/\sqrt{km}}$. The link is a single span of $205~\mathrm{km}$. No laser phase noise is added at this point. After propagation, the central channel is filtered. The signal undergoes matched filtering, chromatic dispersion compensation, and genie-aided multiple-input-multiple-output (MIMO) channel equalization. Finally, we use a fully-data-aided phase filter with an averaging window of $64$ symbols to compensate for the phase rotation induced by cross-phase modulation (XPM), as described in~\cite{fehenberger2015compensation} and we compute the electric SNR and the achievable information rate (AIR) from the equalized signal. The electric signal-to-noise ratio $\mathrm{SNR_{elec}}$ is the SNR computed from the constellation at the end of the DSP chain at the receiver side (i.e., at the input of the decision circuit) and the AIR is computed as Eq. (10) in~\cite{Bocherer14}. For each launch power, five different channel realizations are simulated. 

In the left part of Fig.~\ref{fig:205km_1_span_nonlinear}, we show the electric SNR versus the launch power per channel for the five ESS schemes. We see that the 1D-BL, 4D-BL (Linear), and 4D-BL (Nonlinear) ESS show respectively a $1$~dB, $0.9$~dB, and $1.1$~dB increase in SNR at a power of $10$~dBm compared to conventional ESS, and a $1$~dB enhancement in optimal launch power. In the right part of Fig.~\ref{fig:205km_1_span_nonlinear}, we show the results in terms of AIR computed as Eq.~(1) in~\cite{amari2019enumerative} per 4D symbol versus the optical power. The 4D-BL (linear) and (Nonlinear) show the best AIR. At a launch power of $10$~dBm, the bit rate is increased by $0.07$~bit/4D compared to the 1D-BL-ESS, and by $0.19$~bit/4D compared to conventional ESS. 4D-BL-ESS (Nonlinear) also shows $0.13$~bit/4D gain compared to the 1D-BL-ESS in highly non-linear regime. While K-ESS shows the lowest 1D energy variation as can be seen from Table~\ref{tab:results}, its 4D energy variation is higher than those of the BL-ESS schemes, which is compliant with the measured SNR and AIR performance in Fig.~\ref{fig:205km_1_span_nonlinear}.

\subsection{$5\times80$-km multi-span $110$GBaud transmission}

Next, we compare four schemes: conventional, 1D-BL and 4D-BL ESS (Linear and Nonlinear) in a new scenario at a higher baud rate and over a multi-span link configuration for different shaping block lengths $N$, to assess the scaling of the linear and non-linear gains. We consider a $110$~GBaud transmission over $5\times80$~km SSMF link, $5$ WDM channels with a $112.5$~GHz channel spacing. We also consider a laser linewidth of $100$~kHz at each of the transmitter side and receiver side. The net rate per channel is $838$~Gbit/s ($880$~Gbit/s with $5\%$ pilot overhead). We compensate the added phase noise using a pilot-aided phase estimation~\cite{neves2021joint} with a $5\%$ overhead. The other DSP blocks are the same as the ones used in the previous section. We study the impact of the block length $N$ on the performance in terms of electric SNR and AIR, both measured at the optimal launch power of $3$~dBm. 

In Fig.~\ref{fig:400km_5_span_nonlinear}, we show the variation of the electric SNR, the AIR and the EDI with respect to the shaping block length $N=\{60,108,200,300\}$. For each block length, five different channel realizations are simulated. First, we can seen, by comparing the leftmost and the rightmost figures, that the measured SNR is inversely correlated to EDI changes. However, we also notice that EDI only gives a rough prediction of the performance as it fails to capture small differences such as the gap between the performance of 1D-BL-ESS and 4D-BL-ESS (Linear). Second, we can see that, at a short block length $N=60$, all BL-ESS schemes fail to achieve a significant gain in SNR as the non-linear gains from energy limitation are still small compared to the conventional ESS. Third, when the rate loss decreases as the block length increases, and consequently the linear penalty decreases, the energy fluctuations of conventional ESS reduce the SNR while the SNR of the BL-ESS schemes remain stable up to $N=300$. We did not increase the block length beyond $300$ to maintain a low complexity for the distribution matching. Finally, if we assess the AIR gains through the middle figure, we see that a net performance gain of 
BL-ESS schemes with respect to conventional ESS is observed beyond a block length of $200$. However, all the AIR values achieved by the BL-ESS schemes remain lower than the AIR of conventional ESS for $N\leq 200$ due to their higher rate loss. Our explanation of the limited gains of the BL-ESS for longer distances and higher baud rates is the following: despite the optimization of the signal characteristics at the transmitter side for low NLI distortions through the addition of energy constraints, the chromatic dispersion gradually destroys these characteristics during the transmission. Hence, the band-limited ESS can only obtain limited nonlinear gains over long-distance and high baud-rate transmission (gains achieved over the first few spans), and the gain in AIR is also limited due to the higher rate loss of the band-limited ESS.

\section{Conclusion}
We proposed a new design of ESS schemes that constrains energy variations in higher dimensions (four dimensions in this work) and that can be tuned to boost either linear or nonlinear gains. The new scheme achieves higher gains than state-of-the-art ESS schemes and is particularly interesting for short-distance links as the trade-off between the achieved linear and non-linear gains becomes harder to meet for longer distances because of the accumulation of chromatic dispersion that increases again the energy variations of the optical signal. Besides, our simulation results show that windowed 4D metrics of energy variations predict the non-linear performance of ESS schemes better than 1D and 2D metrics; however, their prediction is not yet accurate enough and the search for better 4D metrics is still an interesting problem to solve. 

\section{Acknowledgements} 
This work was supported by Huawei Technologies France.

\bibliography{ACP23}{}

% Generated by IEEEtran.bst, version: 1.14 (2015/08/26)
\begin{thebibliography}{10}
\providecommand{\url}[1]{#1}
\csname url@samestyle\endcsname
\providecommand{\newblock}{\relax}
\providecommand{\bibinfo}[2]{#2}
\providecommand{\BIBentrySTDinterwordspacing}{\spaceskip=0pt\relax}
\providecommand{\BIBentryALTinterwordstretchfactor}{4}
\providecommand{\BIBentryALTinterwordspacing}{\spaceskip=\fontdimen2\font plus
\BIBentryALTinterwordstretchfactor\fontdimen3\font minus
  \fontdimen4\font\relax}
\providecommand{\BIBforeignlanguage}[2]{{%
\expandafter\ifx\csname l@#1\endcsname\relax
\typeout{** WARNING: IEEEtran.bst: No hyphenation pattern has been}%
\typeout{** loaded for the language `#1'. Using the pattern for}%
\typeout{** the default language instead.}%
\else
\language=\csname l@#1\endcsname
\fi
#2}}
\providecommand{\BIBdecl}{\relax}
\BIBdecl

\bibitem{gultekin2019enumerative}
Y.~C. G{\"u}ltekin, W.~J. van Houtum, A.~G. Koppelaar, F.~M. Willems, J.~Wim
  \emph{et~al.}, ``{Enumerative sphere shaping for wireless communications with
  short packets},'' \emph{IEEE Transactions on Wireless Communications},
  vol.~19, no.~2, pp. 1098--1112, 2019.

\bibitem{schulte2015constant}
P.~Schulte and G.~B{\"o}cherer, ``{Constant composition distribution
  matching},'' \emph{IEEE Transactions on Information Theory}, vol.~62, no.~1,
  pp. 430--434, 2015.

\bibitem{fehenberger2019analysis}
T.~Fehenberger, D.~S. Millar, T.~Koike-Akino, K.~Kojima, K.~Parsons, and
  H.~Griesser, ``Analysis of nonlinear fiber interactions for finite-length
  constant-composition sequences,'' \emph{Journal of Lightwave Technology},
  vol.~38, no.~2, pp. 457--465, 2019.

\bibitem{gultekin2021kurtosis}
Y.~C. G{\"u}ltekin, A.~Alvarado, O.~Vassilieva, I.~Kim, P.~Palacharla, C.~M.
  Okonkwo, and F.~M. Willems, ``{Kurtosis-limited sphere shaping for nonlinear
  interference noise reduction in optical channels},'' \emph{Journal of
  Lightwave Technology}, vol.~40, no.~1, pp. 101--112, 2021.

\bibitem{gultekin2022mitigating}
{G{\"u}ltekin, Yunus Can and Alvarado, Alex and Vassilieva, Olga and Kim,
  Inwoong and Palacharla, Paparao and Okonkwo, Chigo M and Willems, Frans MJ},
  ``Mitigating nonlinear interference by limiting energy variations in sphere
  shaping,'' in \emph{Optical Fiber Communication Conference}.\hskip 1em plus
  0.5em minus 0.4em\relax Optica Publishing Group, 2022, pp. Th3F--2.

\bibitem{wu2022list}
K.~Wu, G.~Liga, A.~Sheikh, Y.~C. G{\"u}ltekin, F.~M. Willems, and A.~Alvarado,
  ``{List-encoding CCDM: A nonlinearity-tolerant shaper aided by energy
  dispersion index},'' \emph{Journal of Lightwave Technology}, vol.~40, no.~4,
  pp. 1064--1071, 2022.

\bibitem{skvortcov2021huffman}
P.~Skvortcov, I.~Phillips, W.~Forysiak, T.~Koike-Akino, K.~Kojima, K.~Parsons,
  and D.~S. Millar, ``Huffman-coded sphere shaping for extended-reach
  single-span links,'' \emph{IEEE Journal of Selected Topics in Quantum
  Electronics}, vol.~27, no.~3, pp. 1--15, 2021.

\bibitem{Schmalen18}
L.~Schmalen, ``Probabilistic constellation shaping: Challenges and
  opportunities for forward error correction,'' in \emph{2018 Optical Fiber
  Communications Conference and Exposition (OFC)}, 2018, pp. 1--3.

\bibitem{wu2021temporal}
K.~Wu, G.~Liga, A.~Sheikh, F.~M. Willems, and A.~Alvarado, ``{Temporal energy
  analysis of symbol sequences for fiber nonlinear interference modelling via
  energy dispersion index},'' \emph{Journal of Lightwave Technology}, vol.~39,
  no.~18, pp. 5766--5782, 2021.

\bibitem{cho2022shaping}
J.~Cho, X.~Chen, G.~Raybon, D.~Che, E.~Burrows, S.~Olsson, and R.~Tkach,
  ``{Shaping lightwaves in time and frequency for optical fiber
  communication},'' \emph{Nature communications}, vol.~13, no.~1, p. 785, 2022.

\bibitem{gultekin2022temporal}
Y.~C. G{\"u}ltekin, K.~Wu, and A.~Alvarado, ``Temporal properties of
  enumerative shaping: Autocorrelation and energy dispersion index,'' in
  \emph{Signal Processing in Photonic Communications}.\hskip 1em plus 0.5em
  minus 0.4em\relax Optica Publishing Group, 2022, pp. SpM2I--3.

\bibitem{fehenberger2015compensation}
T.~Fehenberger, M.~P. Yankov, L.~Barletta, and N.~Hanik, ``Compensation of xpm
  interference by blind tracking of the nonlinear phase in wdm systems with qam
  input,'' in \emph{2015 European Conference on Optical Communication
  (ECOC)}.\hskip 1em plus 0.5em minus 0.4em\relax IEEE, 2015, pp. 1--3.

\bibitem{Bocherer14}
G.~Böcherer, ``{Probabilistic signal shaping for bit-metric decoding},'' in
  \emph{2014 IEEE International Symposium on Information Theory}, 2014, pp.
  431--435.

\bibitem{amari2019enumerative}
A.~Amari, S.~Goossens, Y.~C. G{\"u}ltekin, O.~Vassilieva, I.~Kim, T.~Ikeuchi,
  C.~Okonkwo, F.~M. Willems, and A.~Alvarado, ``{Enumerative sphere shaping for
  rate adaptation and reach increase in WDM transmission systems},'' in
  \emph{45th European Conference on Optical Communication (ECOC 2019)}.\hskip
  1em plus 0.5em minus 0.4em\relax IET, 2019, pp. 1--4.

\bibitem{neves2021joint}
M.~S. Neves, A.~Carena, A.~Nespola, P.~P. Monteiro, and F.~P. Guiomar, ``Joint
  carrier-phase estimation for digital subcarrier multiplexing systems with
  symbol-rate optimization,'' \emph{Journal of Lightwave Technology}, vol.~39,
  no.~20, pp. 6403--6412, 2021.

\end{thebibliography}
\bibliographystyle{IEEEtran}

\end{document}